\RequirePackage{lineno}

\documentclass[aps, twocolumn, prl, superscriptaddress, groupedaddress]{revtex4}
\usepackage{graphicx} 
\usepackage{dcolumn} 
\usepackage{bm} 
\usepackage{amssymb} 
\usepackage{xcolor}%
\usepackage{enumitem}
\usepackage{amsmath, amsfonts}%
\usepackage{scrextend}
\usepackage{ulem}

\newcommand{\appropto}{\mathrel{\vcenter{ \offinterlineskip\halign{\hfil$##$\cr \propto\cr\noalign{\kern2pt}\sim\cr\noalign{\kern-2pt}}}}}

\makeatletter
\renewcommand{\@makefntext}[1]%
{\noindent\makebox[0pt][r]{\textsuperscript{\@thefnmark}\,}#1}
\makeatother

\begin{document}
    \title{Environmental cosmic acceleration from a phase transition in the dark
    sector}

    \author{{\O}.~Christiansen}
    \affiliation{ {Institute of Theoretical Astrophysics}, {University of Oslo}, {{Sem Sælands vei 13}, {0371 Oslo}, {Norway}} }
    \affiliation{ {CEICO, Institute of Physics of the Czech Academy of Sciences, Na Slovance 1999/2, 182 00, Prague 8, Czechia}}
    \author{F.~Hassani}
    \author{D. F.~Mota}
    \affiliation{ {Institute of Theoretical Astrophysics}, {University of Oslo}, {{Sem Sælands vei 13}, {0371 Oslo}, {Norway}} }

    \date{\today}
    \begin{abstract}
        \noindent
        A new degravitation mechanism within the framework of scalar tensor gravity
        is postulated {and included by prescription}. The mechanism eliminates
        all constant contributions from the potential to the Friedmann equation,
        leaving only the kinematic and the dynamic terms of the potential to
        drive cosmic acceleration. We explore a scenario involving a density-triggered
        phase transition in the late-time universe, and argue that the resulting
        effective energy density and equation of state parameter can explain
        late-time cosmology when extrapolated to a region of the parameter space.
    \end{abstract}
    \maketitle
        \noindent
        Cosmological data indicate that the expansion of the universe is currently
        accelerating \cite{planck_planck_2020}. The simplest way to {explain}
        this is a fine-tuned cosmological constant, which is unexplained by the Standard
        Model of particle physics \cite{weinberg_cosmological_1989}. Moreover, tensions
        in recent observational evidence, if not due to systematics, seem to require
        a more phenomenologically complex ingredient than a constant \citep{abdalla_cosmology_2022}.
        Observations also suggest tensions for the dark sector at smaller scales
        \citep{bullock_small-scale_2017}, suggesting the need for more detailed modelling
        of the entire dark sector. This has motivated studies of extensions to
        the Standard Model of cosmology and general relativity.

        Here we present results from a novel explanation of the late-time
        accelerating expansion caused by a modified version of the symmetron
        \citep{hinterbichler_symmetron_2010,hinterbichler_symmetron_2011}. The
        symmetron is a real scalar field with an environment-dependent mass term
        using the Anderson-Higgs mechanism; in low-density environments, its
        vacuum expectation value (VEV) drifts away from the origin, causing a phase
        of out-of-equilibrium transition for the scalar field. {The environmental dependence comes from the scalar's non-minimal coupling to the matter sector, the type of which generically occurs in Horndeski models of modified gravity \cite{horndeski_50_2024}, of which the symmetron is a subset.}
        The VEV also degenerates and takes on a non-trivial topology, leading to
        the appearance of domain walls at cosmological scales. The resulting model
        has a wide range of phenomenological consequences, which individually
        may seem to be able to account for the cosmological observations and
        alleviate tensions
        \cite{hinterbichler_symmetron_2011,burrage_radial_2017,burrage_symmetron_2019,naik_dark_2022,kading_lensing_2023,christiansen_asimulation_2024,christiansen_gravitational_2024}.
        Whether the symmetron, or some variant of it, can simultaneously account
        for the cosmological tensions remains an open question.

        {Phase-transitions as a cause for cosmic acceleration have been considered in the past, in e.g. \cite{avelino_frustrated_2006}, where a frustrated defect network forms as the result of a phase transition in the early universe, showing that the walls' curvature has to be very small. This gives an equation of state parameter $\omega\geq-2/3$. In \cite{bassett_late-time_2002}, the authors agnostically consider a sudden transition in the equation of state parameter of the accelerator field, going from an initial $\omega_{0}= 0$ to some freely chosen $\omega_{f}$ at the cosmological redshift $z_{t}$. Combining CMB data with LSS and supernova data, they indicate parameters of the phase transition of $(z_{t},\omega_{f})\sim (1.5,-1)$, occurring in the late-time universe. The latter is similar to the situation that we will consider, though we will be a step less agnostic and try to realise this situation in a specific variant of the symmetron model. }

        The classical symmetron contains a cosmological
        constant that is included to drive the late-time accelerating expansion \citep{hinterbichler_symmetron_2011},
        which in turn is fine-tuned. In the following, we {add a prescription that we take as representative of some more fundamental model containing a degravitation mechanism. The prescription}
        continuously removes constant contributions {as sources to the Einstein equations (which is what we will mean by degravitation here)};
        instead we try to account for {the} cosmic acceleration through the
        dynamic behaviour of the scalar field.

        Recent advances {in} cosmological simulations of the symmetron \citep{christiansen_asevolution_2023,christiansen_asimulation_2024}
        allow the comprehensive treatment of the scalar field's energy contribution
        for some parameter ranges, that we present the results for here. By extrapolating
        our results, we indicate a region of the model parameter space that is
        interesting with regards to producing the late-time cosmic expansion{, though we still appear to require fine-tuning of the Lagrangian parameters.}

        The field equations of the symmetron $\phi\in \mathbb{R}$ on a flat cosmological
        background is \citep{hinterbichler_symmetron_2011}
            \begin{align}
                \label{eq:eom}\ddot \phi - \frac{1}{a^{2}}\nabla^{2}\phi + 3H\dot \phi = V_{\mathrm{eff},\phi}\equiv V_{,\phi}- \frac{A_{,\phi}(\phi)}{A(\phi)}T,
            \end{align}
        where $a$ is the scale factor, $H$ is the Hubble function, $T$ is the
        trace of the stress-energy tensor of the matter sector, dotted quantities
        are differentiated with respect to cosmic time, $V_{\mathrm{eff}}$ is
        the effective scalar field potential and $V$ is the potential
            \begin{align}
                \label{eq:phipot}V = -\frac{1}{2}\mu^{2}\phi^{2}+ \frac{\lambda}{4}\phi^{4},
            \end{align}
        with mass $\mu$ and self-coupling strength $\lambda$. In the non-relativistic
        limit, $T\sim-\rho_{m}$ is the energy density of matter. $A(\phi)\equiv 1
        + \frac{1}{2}\left(\frac{\phi}{M}\right)^{2}$ is the conformal factor,
        with the conformal coupling strength $M$. {Both the conformal factor $A(\phi )$ and the potential $V(\phi)$ can be viewed as Taylor expansions of a more fundamental model, in the event where their smallness parameters $\epsilon_{A}\sim\left(\phi/M\right)^{2}$ and $\epsilon_{V}\sim\left(\phi/M_{\mathrm{pl}}\right)^{2}$ are small.}
        {The conformal coupling} gives a term for the effective
        potential which provides a stable minimum at the origin as long as
        $\rho_{m}>\rho_{*}=\mu^{2}M^{2}$. {As} $\rho_{m}$ becomes smaller, the
        effective potential becomes the Mexican hat potential, and develops two stable
        minima $\tilde v$ (VEVs) at non-zero field values. The VEVs are {found}
        locally (at time $t$ and position $x$) by minimising $V_{\mathrm{eff}}$
        and are given by
            \begin{align}
                \tilde v_{\pm}(t,x) = \pm\frac{\mu}{\sqrt{\lambda}}\sqrt{ 1 + \frac{ T(t,x)}{\rho_{*}} }.
            \end{align}
        We define the true VEV $v_{0}\equiv \tilde{v}_{+}(T=0)$. The potential $V$,
        equation \eqref{eq:phipot}, in the true vacuum can be related to
        phenomenological quantities \citep{christiansen_asevolution_2023}
            \begin{align}
                \label{eq:potentialAmplitude}V_{0}= \frac{\lambda}{4 }v_{0}^{4}= \frac{9 \beta_{*}^{2}L_{C}^{2}}{2 a_{*}^{6}}H_{0}^{4}\Omega_{m, 0}^{2},
            \end{align}
        where $a_{*}=(1+z_{*})^{-1}$ is the scale factor of the phase transition
        in a homogeneous universe, $\beta_{*}$ is the relative strength of the
        gravitational force in the true VEV $v_{0}$, and $L_{C}$ is the Compton wavelength
        related to the mass $\mu$ as $L_{C}=\frac{1}{\sqrt{2}\mu}$. $H_{0}$ is
        the current time Hubble parameter and $\Omega_{m, 0}$ is the current
        time energy fraction of matter.

        For nonlinear fields such as the symmetron, that may develop large spatial
        gradients due to screening and topological defects, the gradients may
        have a significant effect on the background evolution. It has been shown
        in \cite{christiansen_asimulation_2024} that inhomogeneities in the density
        field cause the field to undergo phase transition much earlier than it would
        on the background. To include this effect into the vacuum, we define the
        generalised potential $V_{\mathrm{gen}}$ whose derivative $V_{\mathrm{gen},\phi}
        \equiv V_{\mathrm{eff},\phi}+\frac{1}{a^{2}}\nabla^{2}\phi$ includes the
        Laplacian term from equation \eqref{eq:eom}. The generalised VEV $v$ is found
        by minimising $V_{\mathrm{gen}}$, i.e. solving
            \begin{align}
                \label{eq:quasistatic}\nabla^{2}v^{2}= -a^{2}\partial_{v}V_{\mathrm{eff}}(v),
            \end{align}
        and is equivalent to finding the symmetron's quasistatic limit (equation
        \eqref{eq:eom} with $\dot\phi ,\ddot\phi \rightarrow 0$).

        In the {usual} approach, such as for quintessence dark energy, the potential
        contains a constant term, $\langle V(\phi=0) \rangle$, which contributes
        to the Friedmann equation, and is the driver of cosmic acceleration. To get
        a constant term that can account for cosmic acceleration requires a high
        degree of fine-tuning (this is the smallness problem of the cosmological
        constant). {We therefore propose} a model where the potential
        contribution to the Friedmann equation is$^{1}$\footnotetext[1]{Using
        $v_{\pm}$ instead, without a prescription for smoothing of $T$, we would
        have contributions from arbitrarily small underdense regions.}
            \begin{align}
                \label{eq:NewPotential}\Omega_{V}(\phi,v) = \frac{\lambda}{4 \rho_{c}}\left\langle\left( \phi^{2}- v^{2}\right)^{2}\right\rangle,
            \end{align}
        which automatically removes all vacuum contributions, and {eliminates} the
        need to fine-tune the cosmological constant{, if a similar prescription can be thought to act on all of the standard model fields}.
        The full contribution to the Friedmann equation {includes} the kinetic
        and gradient terms that are expanded around the VEV,
            \begin{align}
                \label{eq:expandedEnergy}\Omega_{\phi}= \Omega_{V}+ \frac{1}{2\rho_{c}}\langle(\dot{\phi}-\dot v)^{2}+ \partial_{i}(\phi-v) \partial^{i}(\phi-v))\rangle,
            \end{align}
        and we will refer to the last two terms as $\Omega_{\rm{kin}}$ and $\Omega
        _{\rm{grad}}$ respectively. $\rho_{c}(a)$ is the critical density of the
        Universe and the brackets mean that we take the spatial average of their
        argument, because the Friedmann equation is evaluated on the background.
        \\ Since $v$ is the quasistatic solution of the field equation, $\Omega_{V}$
        has contributions from the dynamic parts of the field. In the absence of
        large inhomogeneities in the scalar field (when $\nabla^{2}\phi \longrightarrow
        0$, and $V_{{\rm gen}, \phi}= V_{{\rm eff}, \phi}$), then
        $v\rightarrow \tilde v_{\pm}$, and we recover the classical symmetron energy
        contribution, with an additional term proportional to $v_{\pm}^{4}$. {In the late universe, large inhomogeneities in the symmetron field are sourced by the inhomogeneous matter component that it is coupled to, or caused by the formation of topological defects in the field.}

        We define the dynamical and topological contributions {to the equation \eqref{eq:NewPotential}}
        as $\Omega_{V}=\Omega_{\rm{dyn}}+\Omega_{\rm{top}}$. If the dynamic field,
        $\phi$, develops topological defects through a phase transition, then
        there is a question of how to solve the VEV $v$ in equation \eqref{eq:quasistatic}.
        If we solve it using the same vacuum manifold as is obtained {in} the
        dynamical solution, by which we mean that we are solving the VEV around the
        same topological defects, then we remove the contribution of topological
        defects in equation \eqref{eq:NewPotential}. Alternatively, we can solve
        the VEV using a trivial vacuum manifold (i.e. all {fields with} the same
        positive sign minimum and {no} defects). In this case, the topological
        defects {that develop} in the field $\phi$ will introduce large
        gradients for $\phi$ that are absent in the VEV $v$ and will therefore
        give large contributions in equation \eqref{eq:NewPotential}.
        Considering both solutions for the VEV allows us to separate the effect
        of topology $\Omega_{\rm{top}}$.

        \begin{figure*}
            \centering
            \includegraphics[width=0.9\linewidth]{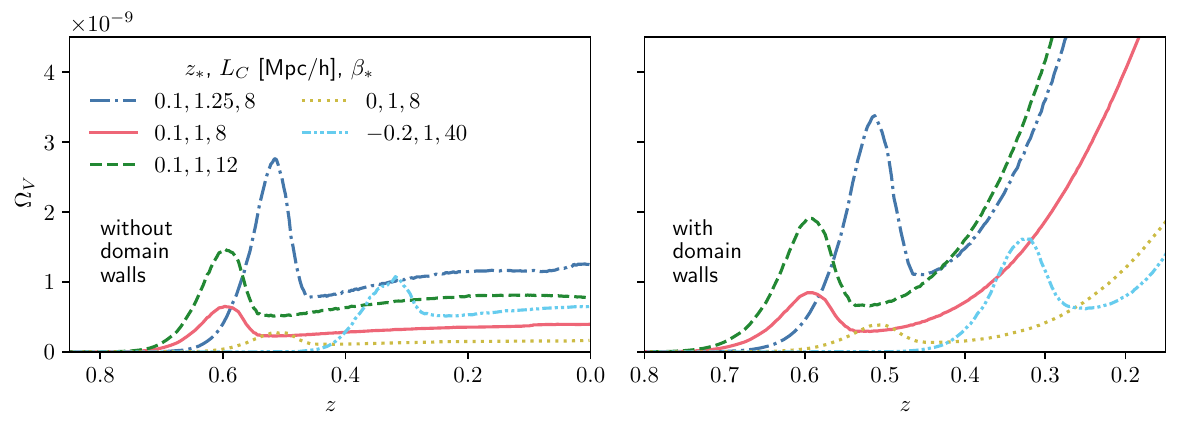}
            \caption{ Energy density contribution from the scalar field potential,
            normalised to the critical density of the Universe ($\Omega=\rho/\rho
            _{c}$), as a function of redshift $z$. The left panel shows the case
            where the domain wall contribution is removed. Right panel includes the
            contribution of the domain walls.}
            \label{fig:potentialenergy}
        \end{figure*}
        \begin{figure*}
            \centering
            \includegraphics[width=0.9\linewidth]{
                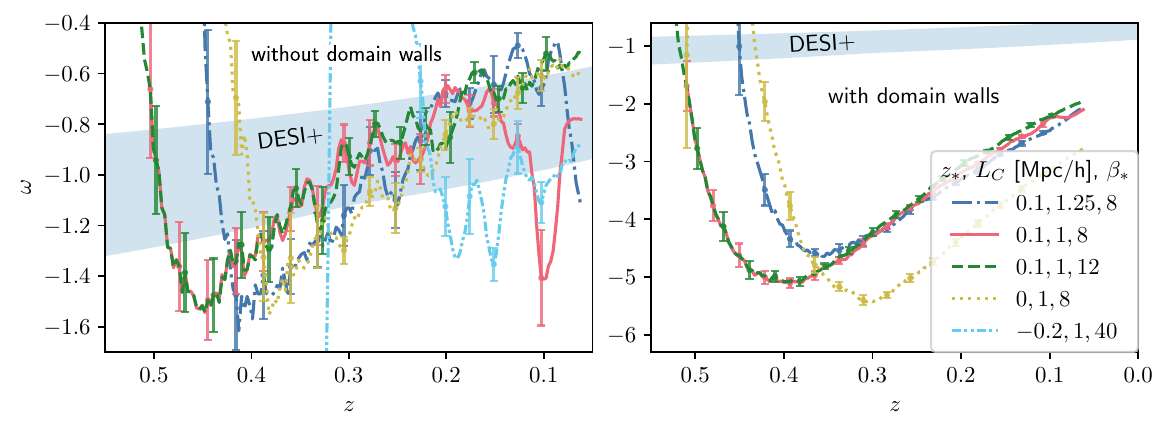
            }
            \caption{Effective equation of state
            parameters as a function of cosmological redshift $z$. The kinematic
            and gradient contributions \eqref{eq:expandedEnergy} are neglected
            here. The left panel shows the case where the domain wall contribution
            is removed. The right panel includes the domain wall contribution. Error
            bars show 95 \% confidence intervals due to the intrinsic scatter in
            the data. The band labelled DESI+ demonstrates the 68\% preferred
            region for the flat $\omega_{0}\omega_{a}$CDM model, presented in
            \cite{desi_collaboration_desi_2024}.}
            \label{fig:equationOfState}
        \end{figure*}

        {We simulate the matter and scalar field evolution through the phase transition from initially small amplitude perturbations $\phi/v_{0}\sim 10^{-20}$ as in \cite{christiansen_asimulation_2024}. Now, we additionally keep track of the vacuum field $v$, by solving the equation \eqref{eq:quasistatic}, considering both options for the vacuum topology as mentioned in the previous paragraph. We conveniently express the model parameters ($\mu, M, \lambda$) in terms of phenomenological parameters ($z_{*}, L_{C}, \beta_{*}$), as in \cite{christiansen_asevolution_2023}, where $z_{*}$ is the redshift of the phase transition, $L_{C}$ is the interaction length scale and $\beta_{*}$ is the relative strength of the gravitational force in the true vacuum $v_{0}=\mu/\sqrt{\lambda}$. The parameters are chosen in the same range as in \cite{christiansen_asimulation_2024} ($z_{*}, L_{C}, \beta_{*}\,\sim\,0,\,1\,\mathrm{Mpc/h},\,10$), where the simulations' convergence and consistency has been studied in detail, and we are resolving the relevant scales. This corresponds to Lagrangian parameters $\mu$, $M$, $\lambda$ of around $10^{-30}\,\mathrm{eV}$, $10^{-4}\,M_{\mathrm{pl}}$ and $10^{-104}$ respectively. For these choices, the smallness parameters $\epsilon_{A},\epsilon_{V}\sim 10^{-5}, 10^{-12}$. $\epsilon_{A}\ll1$ is required by the simulation scheme, see \cite{christiansen_asevolution_2023}. We record the resultant volume averaged energy contributions according to equations \eqref{eq:NewPotential} and \eqref{eq:expandedEnergy}. }

        In figure \ref{fig:potentialenergy}{,} we plot the volume average,
        equation \eqref{eq:NewPotential}, for some different parameter choices. We
        see that 1) In all cases, there is an initial spike in the energy density
        at the start of the phase transition; 2) for this choice of parameters,
        the energy scale is of {the} order {of} $\Omega_{V}(z=0)\sim 10^{-9}$. {To}
        obtain the energy scale $\mathcal{O}(1)$ required for the accelerating expansion,
        we have to extrapolate in the model parameters; 3) the effect of
        changing $\beta_{*}$ is very well approximated$^{2}$\footnotetext[2]{We
        have {generally found} a very small effect of {varying} $\beta$ on the
        scalar field configuration \citep{christiansen_asimulation_2024}.} by
        the change in amplitude given by the equation \eqref{eq:potentialAmplitude},
        which means that $\langle V\rangle\appropto \beta_{*}^{2}$. Considering all
        of the parameter scalings, we find
        $V\appropto \frac{\beta_{*}^{2}L_{C}^{5}}{ a_{*}^{9}}$; and 4) in all
        cases where the domain wall contribution (topology) is included in the
        energy density, there is a rapidly increasing potential energy in the later
        stages of the phase transition.

        We define the effective equation of state through the energy contribution
        of the scalar field \citep{christiansen_asevolution_2023}, $\rho$, as
            \begin{align}
                \label{eq:eos}\omega = \frac{1}{3}\frac{\mathrm{d} \log {\rho}}{\mathrm{d} \log a}-1.
            \end{align}
            
        Figure \ref{fig:equationOfState} shows the effective equation of state
        parameter carrying the energy of the potential in equation \eqref{eq:NewPotential}.
        We are neglecting the kinematic and gradient contributions
        $\Omega_{\rm{kin}},\Omega_{\rm{grad}}$ for now. In all cases in figure \ref{fig:equationOfState}
        where we remove the effect of the domain walls (topology), the effective
        equation of state parameter is centred around $- 1$. There is a slight drift
        from a phantom regime ($\omega<-1$) to larger values and some oscillation
        around this drift. An earlier phase transition seems to reduce the slope
        of the drift. In the case where the effect of domain walls (topology) is
        included, the topological defects seem to push the equation of state
        parameter towards the phantom regime. Usually in literature, having a `phantom
        crossing' is considered problematic\citep{vikman_can_2005}, but their argument
        does not apply in our case due to the degravitation mechanism and matter
        coupling.

        Finally, there are the kinetic and gradient terms, $\Omega_{\rm{kin}}$ and
        $\Omega_{\rm{grad}}$, which we also expect to contribute to the
        Friedmann equation. Our results show that they both peak early {in} the phase
        transition, after which the gradient contribution decays rapidly and
        $\Omega_{\rm{kin}}$ dominates {between} the two. {We also} find that the
        vacuum $v$ {generally} evolves slowly $|\dot v| \ll |\dot \phi|$ and so $\Omega
        _{\rm{kin}}$ is the contribution of the scalar field kinetic energy $\dot
        \phi^{2}/2$ coming from the oscillations of the field. We find {that}
        the kinetic term {is} larger than the potential contribution, equation \eqref{eq:NewPotential},
        for all of the models. The kinetic energy decays as
        $\rho_{\rm{kin}}\sim a^{-2}$.

        Comparing different simulations, we see no effect of parameter variation
        on the relative amplitude of the kinetic and potential components.
        However, we do see a smaller injection of kinetic energy at the start of
        the phase transition in the case of stronger screening (or larger
        $a_{*}$). The height of the peak compares to the plateau as
        $\Omega_{\mathrm{kin}}^{\mathrm{(peak)}}/\Omega_{\mathrm{kin}}\appropto \sqrt{\frac{L_{C}}{a_{*}}}$.
        This factor is $\Omega_{\mathrm{kin}}^{\mathrm{(peak)}}/\Omega_{\mathrm{kin}}
        \approx 2.7$ for the fiducial simulation ($z_{*},\,L_{C},\,\beta_{*}$\,=\,0.1,\,1,\,8).
        While for the same parameters, today ($z=0$),
        $\Omega_{\mathrm{kin}}/\Omega_{V}\sim 4$.

        From the above, we can see that in order to dynamically {generate} an energy
        density that mimics the cosmological constant by the mechanism proposed here,
        the following problems must be solved by moving to an appropriate region
        of the parameter space
        \begin{enumerate}[label=\Alph{*}]
            \itemsep-.3em

            \item The initial spike of energy density occurring at the start of the
                phase transition.

            \item The late rise in energy owing to a growing contribution of the
                domain walls to the energy budget.

            \item The large amplitude of the kinetic energy relative to the potential
                $\Omega_{\rm{kin}}/\Omega_{V}$.

            \item The small amplitude of the energy density $\Omega_{V}$.
        \end{enumerate}

        Therefore, we note the following part of the symmetron parameter space
        as interesting with regards to issues A-D: small critical energy densities
        $\rho_{*}$ (equivalently large $a_{*}> 1$) and small interaction length scales
        $L_{C}$. Finally the force strength $\beta_{*}$ is chosen to obtain the
        correct energy density amplitude.

        By having small critical energy densities $\rho_{*}$ we ensure that only
        the least dense regions, which are mostly disjunct up {to} the {present}
        time undergo the phase transition. {Since} the domains are mostly
        disjunct, we expect a suppression of the final {increase} in energy density
        (B) that is caused by the formation of domain walls. $a_{*}>1$
        increasingly depends on nonlinearities in the density field to undergo
        collapse before the current time, {so} the scale $L_{C}$ is chosen {to make}
        the field interact with these small-scale nonlinearities. In this regime,
        we expect an almost quasistatic transition, since the timescale of the
        scalar field is smaller and the spatial scale {resolves} the filaments. We
        therefore expect a suppression of the initial spike (A) that we understand
        to be caused by an initial period of out-of-equilibrium dynamics for the
        scalar.

        We do not see a clear effect of varying the parameters on the relative amplitude
        between {the} kinetic and potential energies, but we {have} {found}
        nonlinearities in the dependence on $z_{*}$ and $L_{C}$, that make the extrapolation
        {difficult}. As the initial spike (A) is the first injection of kinetic
        energy into the field, we expect a suppression of the relative amplitude
        of the kinetic energy (C). The problem (C), may therefore be resolved in
        the interesting limit of small $L_{C},z_{*}$.

        To be explicit, we make the following estimates: To remove the initial spike
        (A), we assume that it is sufficient that the spatial scale of the
        symmetron is of order $L_{C}\lesssim \mathcal{O(\rm{kpc})}$. For such a choice,
        considering the inferred scaling of
        $a_{\rm{SSB}}\appropto L_{C}^{0.22}a_{*}^{0.59}$, keeping $z_{\rm{SSB}}\sim
        0.6$, we set $a_{*}\sim 12$. This is motivated {by} cosmological data{,}
        where {the} cosmic acceleration happens around
        $0.4 \lesssim z \lesssim 1$ \citep{planck_planck_2020}. Finally, {to obtain}
        the correct energy amplitude (D), given the reported scaling of $V$, we set
        $\beta_{*}= 10^{18}$. Although this may seem unreasonably large compared
        to the values of $\beta_{*}\sim \mathcal{O}(1)$ typically considered in
        the literature, we point out that the screening is also set very large,
        so that the force operates {only} inside of the least dense voids{,} where
        {the} clustering constraints are at their worst and the effect of a fifth
        force is minimal. Moreover, this force strength is still $10^{7}$ times weaker
        than the weak force and $10^{18}$ times weaker than electromagnetism. 
        
        {The resultant model is one where the nearly quasistatic field's
        displacement from the drifting and environmentally dependent vacuum, is
        the cause of cosmic acceleration. }
        Despite the field being strongly screened, we expect observables such as
        a modification of the density profile within large cosmic voids, which is
        expected to give clear signatures in observables such as the Integrated Sachs-Wolfe
        (ISW) effect \citep{christiansen_asimulation_2024,kovacs_imprint_2017,kovacs_part_2018,kovacs_more_2019}.
        We expect upcoming constraints on the equation of state parameter from among
        other the continuation of the DESI survey \citep{desi_collaboration_desi_2024}.
        And the oscillating energy density and topological defects can leave observable
        imprints in gravitational waves \cite{afzal_nanograv_2023,christiansen_gravitational_2024}.

        The appeal of this mechanism is manifold:
        \begin{enumerate}[topsep=0.2em]
            \itemsep-.3em

            \item The nearly quasistatic evolution of the field {as a generator of cosmic acceleration}
                addresses the smallness problem of the cosmological constant by
                suppressing the energy density contribution to the Friedmann
                equation relative to that of the original potential \eqref{eq:phipot}

            \item Since the phase transition is triggered by nonlinearities in the
                overdensity field, the mechanism partially solves the
                coincidence problem by making `now', { when nonlinearities grow large,}
                (i.e. $z\lesssim 1$) a generic time for the accelerated expansion

            \item The addition of a fifth force at a strength scale midway between
                electromagnetism and gravity improves the hierarchy problem of why
                gravity is uniquely weak, and provides a more continuous range
                of force scales among the five forces

            \item In terms of minimality, the symmetron has been considered as a
                dark matter candidate for similar parameter choices for the
                interaction scale $L_{C}\sim$kpc \citep{burrage_radial_2017,burrage_symmetron_2019,kading_lensing_2023},
                in which case it behaves as a fuzzy dark matter with an extended
                mass spectrum that depends on the environment

            \item The relatively large energy scale of the degravitated potential
                $V_{0}$ and the phantom behaviour caused by the formation of topological
                defects may allow the connection of the low energy, late-time
                acceleration with the higher energy scale expansion during
                inflation {\citep{vilenkin_topological_1994}} in a cyclic
                universe scenario. In this scenario, the symmetron has the
                additional {advantage} of being symmetry restored during
                reheating. Looking at the degravitated potential $V_{0}$ from {the}
                equation \eqref{eq:potentialAmplitude}, we see that it has a scale
               { $V_{0}\sim 10^{29}\cdot\rho_{c,0}\sim 10^{39}$ GeV/m$^{3}=10^{19}\left(\mathrm{eV}\right)^4$}, which
                can be made available in the topological defects or upon symmetry
                restoration.
        \end{enumerate}

        { However, in spite of these optimistic remarks,
        a change of perspective from the consideration of the
        phenomenological parameters $(L_{C},a_{*},\beta)$, to
        the Lagrangian ones $(\mu,M,\lambda)$, reveals caveats.
        While $L_{C}\sim$kpc can make the symmetron an interesting
        fuzzy dark matter candidate, the mass $\mu$ is then of order $10^{-27}$ eV, lighter than the typical standard model mass scale by 27 orders of magnitude. Meanwhile $a_{*}=12$ corresponds to conformal coupling parameters $M\sim 10^{20}$ eV, or in terms of the Planck mass $M/M_{\mathrm{pl}}\sim 10^{-8}$. Finally, the force strength $\beta_{*}=10^{18}$, together with the other two parameter choices, results in the self-coupling strength $\lambda\sim 10^{-114}$. From naturalness, we expect dimensionless numbers $\lambda\sim\mathcal{O}(1)$. While, the nature of the fine-tuning issue here is different from the cosmological constant, where the prior for a 120 orders of magnitude larger cosmological constant value comes from particle physics predictions, the discrepancy is of similar magnitude. A first-principles understanding to explain the apparent fine tuning remains to be found. Furthermore, the resultant smallness parameters $\epsilon_{A},\epsilon_{V}\sim 10^{20},10^{4}$ both indicate the relevance of higher order operators within this part of the parameter space, a study into which is beyond the scope of the current work. }

        In conclusion, the large discrepancy between the cosmological constant
        expected from quantum field theory and that required to explain the
        {cosmological} acceleration motivates: 1) the introduction of a
        degravitation mechanism of constant term contributions to the Friedmann
        equation and 2) dynamical drivers of the cosmological acceleration. In
        this work we have introduced a model where there is a late-time, environment-dependent
        phase transition in the dark sector whose dynamic energy component drives
        the late-time accelerated expansion. We have studied the evolution and
        parameter scaling of the energy components within the parameter space
        computationally accessible to our simulations. By extrapolation, we
        point to a specific part of the symmetron parameter space, $(L_{C},a_{*},
        \beta_{*})\sim (\rm{kpc},12,10^{18})$, which can account for the late-time
        cosmic acceleration.
        \newline
        \newline
        \noindent
        \small{\textit{Acknowledgements: {The authors are grateful to the anonymous referees for their constructive comments that helped us to improve the manuscript. } {\O}.C. thanks Martin Kunz for discussions.
        This work was co-funded by the European Union and supported by the Czech Ministry of
        Education, Youth and Sports (Project No. FORTE – CZ.02.01.01/00/22\_008/0004632).
        We thank the Research Council of Norway for their support and the resources provided by UNINETT Sigma2 -- the National Infrastructure for High Performance Computing and Data Storage in Norway, as well as the Swiss National Supercomputing Centre for their support under project IDs 301367 and s1051 respectively. }}

\end{document}